\documentclass{article}

\usepackage{amsfonts}
\usepackage{amsmath}
\usepackage{amssymb}
\newtheorem
{proposition}{Proposition}

\begin{document}

\title{Quasiclassical generalized Weierstrass representation and
dispersionless DS equation}
\author{B.G.Konopelchenko \\
Dipartimento di Fisica, Universita del Salento \\
and INFN, Sezione di Lecce, 73100 Lecce, Italy}
\date{}
\maketitle

\begin{abstract}

Quasiclassical generalized Weierstrass representation (GWR) for highly
corrugated surfaces in $\mathbb{R}^{4}$ with slow modulation is proposed.
Integrable deformations of such surfaces are described by the dispersionless
Davey-Stewartson (DS) hierarchy. Quasiclassical GWRs for other four-dimensional
spaces and dispersionless DS system are discussed too.
\end{abstract}

\section{Introduction}

\setcounter{equation}{0}

Classical Weierstrass representation is the basic analytic tool to study
and analyze minimal surfaces both in mathematics and in aplications (see
\textit{e.g.} [1-3]). Its generalizations to generic surfaces conformally
immersed into the three and four dimensional spaces have been proposed
recently in [4-8]. These generalized Weierstrass representations (GWRs)
were based on the two-dimensional Dirac equation and they allow
to costruct any analytic
surface in $R^{4}$and $R^{3}$. The hierarchies of the Davey-Stewartson (DS)
and modified Veselov-Novikov(mVN) equations generate the integrable
deformations of surfaces in $R^{4}$ and $R^{3}$ respectively [4,6]. GWRs
provide us with an effective analytic tool to study various problems both
for generic surfaces and special classes of surfaces. GWRs are also quite
useful in numerous applications in applied mathematics, string theory,
membrane theory and other fields of physics (see \textit{e.g.} [9-17]).

The most of the papers on this subject and results obtained have been
concerned to a smooth case. On the other hand, irregular, corrugated
surfaces also have attracted interest in various fields from the applied
physics, technology to the pure mathematics (see \textit{e.g.} [17-25]).

In the present paper we propose a Weierstrass type representation for highly
corrugated surfaces with a slow modulation in the four and three dimensional
Euclidean spaces. It is the quasiclassical limit of the generalized
Weierstrass representation for surfaces in $R^{4}$ and $\mathbb{R}^{3}$
introduced in [4-6]. The quasiclassical GWR is based on the quasiclassical
limit of the Dirac equation. It allows us to construct surfaces in $R^{4}$
and $\mathbb{R}^{3}$ with highly oscillating (corrugated) profiles and slow
modulations of these oscilations characterized by a small parameter $%
\varepsilon =\frac{l}{L}$ where l and L are typical scales of oscillations
and modulations, respectively. In the lowest order in $\varepsilon $ the
coordinates $X^{j}$ ($j=1,2,3,4$) of such surfaces in $R^{4}$are of the form
\begin{eqnarray}
X^{1}+iX^{2} &=&A(\varepsilon z,\varepsilon \overline{z})\exp \left[ i\frac{%
S_{12}(\varepsilon z,\varepsilon \overline{z})}{\varepsilon }\right] ,
\notag \\
X^{3}+iX^{4} &=&B(\varepsilon z,\varepsilon \overline{z})\exp \left[ i\frac{%
S_{34}(\varepsilon z,\varepsilon \overline{z})}{\varepsilon }\right]
\label{1.1}
\end{eqnarray}%
where $z$ and $\overline{z}$ are the conformal coordinates on a surface, A
and B are some smooth functions and $S_{12}$ and $S_{34}$ are related to
solutions of the eikonal type equation. The corresponding metric and mean
curvature are finite functions of the slow variables $\varepsilon
z,\varepsilon \overline{z}$ while the Gaussian curvature is of the order $%
\varepsilon ^{2}$.

Integrable deformations of such corrugated surfaces in $R^{4}$ and $R^{3}$
are induced by the hierarchy of dispersionless DS-II (dDS) equations and
dispersionless mNV (dmNV) equations . These deformations preserve the
quasiclassical limit of the Willmore functional (Canham-Helfrich bending
energy for membranes or the Polyakov extrinsic action for strings). The
dispersionless limit of the generic DS system is considered too. Generalized
Weierstrass representation for surfaces in $\mathbb{R}^{4}$

\section{Generalized Weierstrass representation for surfaces in $R^{4}$}

\setcounter{equation}{0}

The generalized Weierstrass representation (GWR) for surface in $R^{4}$
proposed in [6] is based on the linear systems (two-dimensional Dirac
equations)
\begin{eqnarray}
\Psi _{1z} &=&p\Phi _{1},  \notag \\
\Phi _{1\overline{z}} &=&-\overline{p}\Psi _{1},
\end{eqnarray}%
\begin{eqnarray}
\Psi _{2z} &=&\overline{p}\Phi _{2,} \notag \\
\Phi _{2\overline{z}} &=&-p\Psi _{2}
\end{eqnarray}%
where $\Psi _{k}$ and $\Phi _{k}$ $(k=1,2)$ are complex -valued functions of
$z,\overline{z}\in \mathbb{C}$ (bar denotes a complex conjugation) and $p(z,%
\overline{z})$ is a complex-valued function. One then defines four
real-valued functions $X^{j}(z,\overline{z}),j=1,2,3,4$ by the formulae
\begin{eqnarray}
X^{1}+iX^{2} &=&\int_{\Gamma }(-\Phi _{1}\Phi _{2}dz^{\prime} +
\Psi _{1}\Psi_{2}d\overline{z^{\prime}}), \\
X^{3}+iX^{4} &=&-\int_{\Gamma }(\Phi _{1}\overline{\Psi }_{2}dz^{\prime}
+\Psi_{1}\overline{\Phi }_{2}d\overline{z^{\prime} })
\end{eqnarray}%
where $\Gamma $ is an arbitrary contour in $\mathbb{C}$.


\begin{proposition} {\textbf{[6]}}
For any function $\ p(z,\overline{z})$ and any
solutions ($\Psi _{k},\Phi _{k}$) of the system (2.1)-(2.2),
the formulae (2.3)-(2.4)
define a conformal immersion of a surface into $\mathbb{R}^{4}$ with
the induced metric
\begin{equation}
ds^{2}=u_{1}u_{2}dzd\overline{z},
\end{equation}%
the Gaussian curvature
\begin{equation}
K=-\frac{4}{u_{1}u_{2}}(\log u_{1}u_{2})_{z\overline{z}},
\end{equation}%
squared mean curvature vector
\begin{equation}
\mathbf{H}^{2}=2\frac{\left| p\right| ^{2}}{u_{1}u_{2}},
\end{equation}%
and the Willmore functional given by
\begin{equation}
W\overset{def}{=}\iint_{G}\mathbf{H}^{2}[ds]=4\iint_{G}\left| p\right|
^{2}dxdy
\end{equation}%
where $u_{k}=\left| \Psi _{k}\right| ^{2}+\left| \Phi _{k}\right| ^{2}$ and $%
z=x+iy$.
Morever, any regular surface in $\mathbb{R}^{4}$ can be constructed via the
GWR (2.1)-(2.4) [6, 15].
\end{proposition}

In the particular case $p=\overline{p,}\Psi _{2}=\pm \Psi _{1},\Phi _{2}=\pm
\Phi _{1}$ one has $X_{z}^{4}=X_{\overline{z}}^{4}=0$ and the formulae
(2.1)-(2.4) define surface in $R^{3}$ [4,5].

Integrable dynamics of surfaces constructed via the GWR (2.1)- (2.4) is
induced by the integrable evolutions of the potential $p(z,\overline{z},t)$
and the functions $\Psi _{k},\Phi _{k}$ with respect the \ deformation
parameters $t_{l}$. \ They are given by the DS -II hierarchy [6,7]. The
simplest example is the DS-II equation
\begin{align*}
&ip_{t_{2}}+p_{zz}+p_{\overline{z}\overline{z}}+(\omega _{1}+\omega _{2})p
=0, \\
& \omega _{1z} =2\left| p\right| _{\overline{z}}^{2},
\qquad \omega _{2\overline{z}}=2\left| p\right| _{z}^{2}
\tag{2.8}
\end{align*}
for which
\begin{align*}
&i\Psi _{1t_{2}}+\Psi _{1\overline{z}\overline{z}}+\omega _{1}\Psi
_{1}+p_{z}\Phi _{1}-p\Phi _{1z} =0, \\
& i\Phi _{1t_{2}}+\overline{p}_{\overline{z}}\Psi _{1}-\overline{p}\Psi _{1%
\overline{z}}-\Phi _{1zz}-\omega _{2}\Phi _{1} =0
\tag{2.9}
\end{align*}
and
\begin{align*}
& -i\Psi _{2t_{2}}+\Psi _{2\overline{z}\overline{z}}+\omega _{1}\Psi _{2}+%
\overline{p}_{z}\Phi _{2}-\overline{p}\Phi _{2z} =0, \\
& -i\Phi _{2t_{2}}+p_{\overline{z}}\Psi _{2}-p\Psi _{2\overline{z}}-\Phi
_{2zz}-\omega _{2}\Phi _{2} =0.
\tag{2.10}
\end{align*}

In the reduction to the three dimesnional case the constraint $p=\overline{p}
$ is compatible only with odd DS-II flows (times $t_{2l+1}$) and the DS
hierarchy is reduced to the mVN hierarchy. The lowest member of this
hierarchy is given by the mVN equation
\begin{align}
&p_{t}+p_{zzz}+p_{\overline{z}\overline{z}\overline{z}}+3\omega p_{z}+3%
\overline{\omega }p_{\overline{z}}+\frac{3}{2}p\omega _{z}+\frac{3}{2}p%
\overline{\omega }_{\overline{z}} =0,  \notag \\ &
\omega _{\overline{z}} =(p^{2})_{z}. \tag{2.11}
\end{align}

The DS equation (2.8) and the whole DS hierarchy are amenable to the inverse
spectral transform method (see \textit{e.g.} [26,27])
and they have a number of
remarkable properties typical for integrable 2+1-dimensional equations.
Integrable dynamics of surfaces in $\mathbb{R}^{4}$ inherits all these
properties [6,7]. One of the remarkable features of such dynamics is that
the Willmore functional W (2.7) remains invariant ($W_{t}=0$) [6,7]. In
virtue of the linearity of the basic problem (2.1) the GWR is quite a useful
tool to study various problems in physics and mathematics (see \textit{e.g.}
[5-17]).

\section{Quasiclassical Weierstrass representation.}

\setcounter{equation}{0}

In this paper we shall consider a class of surfaces in $\mathbb{R}^{4}$
which can be characterized by two scales l and L such that the parameter $%
\varepsilon =\frac{l}{L}\ll 1.$ A simple example of such a surface is
provided by the profile of a slowly modulated wavetrain for which l is a
typical wavelength and L is a typical length of modulation. Theory of such
highly oscillating waves with slow modulations is well developed (see
\textit{e.g.} [28,29]). Following the ideas of this Whitham (or nonlinear
WKB) theory we will study surfaces in $\mathbb{R}^{4}$ for which the
coordinates $X^{1},X^{2},X^{3},X^{4}$ have the form

\begin{equation}
X^{i}(z,\overline{z})=\sum_{n=0}^{\infty }\varepsilon ^{n}F_{n}^{i}\left(
\frac{\overrightarrow{S}(\varepsilon z,\varepsilon \overline{z})}{%
\varepsilon },\varepsilon z,\varepsilon \overline{z}\right) ,\quad
i=1,2,3,4
\end{equation}%
where $\overrightarrow{S}=(S^{1},S^{2},S^{3},S^{4})$ and $F_{n}^{i}$ are
smooth functions of slow variables $\xi =\varepsilon z,\overline{\xi }%
=\varepsilon \overline{z}$ and the small parameter $\varepsilon $ is defined
above. The arguments $\frac{S^{i}}{\varepsilon }$ in $F_{n}^{i}$ describe a
fast variation of a surface while the rest of arguments correspond to slow
modulations.

There are different ways to specify functions $\ F_{n}^{i}$. Here we will
consider one of them induced by the similar quasiclassical (WKB) limit of
the GWR (2.1)-(2.4).

Thus, we begin with the quasiclassical limit of the Dirac equations
(2.1)-(2.2).
Following the discussion of the one-dimensional case by Zakharov [30]
(dispersionless limit of the nonlinear Schrodinger equation), we take
\begin{align*}
&p=\exp \left( \frac{i(S_{1}(\varepsilon z,\varepsilon \overline{z})-%
\widetilde{S}_{1}(\varepsilon z,\varepsilon \overline{z}))}{\varepsilon }%
\right) \sum_{n=0}^{\infty }\varepsilon ^{n}p_{n}(\varepsilon z,\varepsilon
\overline{z}), \tag{3.2} \\
&\Psi_{k}=\exp \left( \frac{iS_{k}(\varepsilon z,\varepsilon
\overline{z})}{\varepsilon }\right) \sum_{n=0}^{\infty }
\varepsilon ^{n}\Psi_{kn}(\varepsilon z,\varepsilon \overline{z}), \tag{3.3} \\
&\Phi _{k}=\exp \left( \frac{i\widetilde{S}_{k}(\varepsilon z,\varepsilon
\overline{z})}{\varepsilon }\right) \sum_{n=0}^{\infty }\varepsilon ^{n}\Phi
_{kn}(\varepsilon z,\varepsilon \overline{z}) \tag{3.4}
\end{align*}%
where $S_{k},\widetilde{S}_{k},\Psi _{kn},\Phi _{kn}$, $k=1,2$ are smooth
functions of slow variables $\xi =\varepsilon z,\overline{\xi }=\varepsilon
\overline{z}$ $,$ $\overline{S}_{k}=S_{k}$ ,$\overline{\widetilde{S}}_{k}=%
\widetilde{S}_{k}$ and $S_{1}+S_{2}=\widetilde{S}_{1}+\widetilde{S}_{2}$.
Properties of the asymptotic expansions of the type (3.1)-(3.4) are
quite well studied [28,29]. Here we restrict ourselfs to the lowest order
terms. In this order one has
\begin{align*}
&\Psi _{1}=\Psi _{10}\exp (\frac{iS_{1}}{\varepsilon }),\Phi _{1}=\Phi
_{10}\exp (\frac{i\widetilde{S}_{1}}{\varepsilon }), \tag{3.5} \\
& \Psi _{2}=\Psi _{20}\exp (\frac{iS_{2}}{\varepsilon }),\Phi _{2}=\Phi
_{20}\exp (\frac{i\widetilde{S}_{2}}{\varepsilon }),\tag{3.6} \\
& p=p_{0}\exp (\frac{i(S_{1}-\widetilde{S}_{1})}{\varepsilon }). \tag{3.7}
\end{align*}
Substituting these expressions into (2.1)-(2.2), one in zero order in
$\varepsilon $ gets the algebraic equations
\begin{equation*}
\left(
\begin{array}{cc}
iS_{1\xi } & -p_{0} \\
\overline{p}_{0} &i\widetilde{S}_{1\overline{\xi }}%
\end{array}%
\right) \left(
\begin{array}{c}
\Psi _{10} \\
\Phi _{10}%
\end{array}%
\right) =0, \qquad \quad
\left(
\begin{array}{cc}
iS_{2\xi } & -\overline{p}_{0} \\
p_{0} & i\widetilde{S}_{2\overline{\xi }}%
\end{array}%
\right) \left(
\begin{array}{c}
\Psi _{20} \\
\Phi _{20}%
\end{array}%
\right) =0.
\tag{3.8}
\end{equation*}

The existence of nontrivial solutions for these systems imply that
$p_{0}$ and $S_{k},\widetilde{S}_{k}$ should obey the equations

\begin{align*}
&
\det \left(
\begin{array}{c}
iS_{1\xi },-p_{0} \\
\overline{p}_{0},i\widetilde{S}_{1\overline{\xi }}%
\end{array}%
\right) =-S_{1\xi }\widetilde{S}_{1\overline{\xi }}+\left| p_{0}\right|
^{2}=0, \tag{3.9}
\\
&
\det \left(
\begin{array}{c}
iS_{2\xi },-\overline{p}_{0} \\
p_{0},i\widetilde{S}_{2\overline{\xi }}%
\end{array}%
\right) =-S_{2\xi }\widetilde{S}_{2\overline{\xi }}+\left| p_{0}\right|
^{2}=0.\tag{3.10}
\end{align*}
Further, using the differential form of (2.3)-(2.4), namely, equations

\begin{align*}
& \left( X^{1}+iX^{2}\right) _{z}=-\Phi _{1}\Phi _{2},
&(X^{1}+iX^{2})_{\overline{z}}=\Psi _{1}\Psi _{2},
\\
& (X^{3}+iX^{4})_{z}=\Phi _{1}\overline{\Psi }_{2},
& (X^{3}+iX^{4})_{\overline{z}}=\Psi _{1}\overline{\Phi }_{2},
\tag{3.11}
\end{align*}
one concludes that in the lowest order in $\varepsilon $ one has
\begin{align*}
&X^{1}+iX^{2} =(X_{0}^{1}+iX_{0}^{2})\exp (\frac{iS_{12}}{\varepsilon }),
\tag{3.12} \\
&X^{3}+iX^{4} =(X_{0}^{3}+iX_{0}^{4})\exp (\frac{iS_{34}}{\varepsilon })
\tag{3.13}
\end{align*}
where
\begin{align*}
&S_{12} =S_{1}+S_{2}=\widetilde{S}_{1}+\widetilde{S}_{2},\tag{3.14} \\
&S_{34} =S_{1}-\widetilde{S}_{2}=\widetilde{S}_{1}-S_{2} \tag{3.15}
\end{align*}
and
\begin{align*}
&X_{0}^{1}+iX_{0}^{2} =-i\frac{\Psi _{10}\Psi _{20}}{(S_{1}+S_{2})_{\xi }}=i%
\frac{\Phi _{10}\Phi _{20}}{(S_{1}+S_{2})_{\overline{\xi }}}, \tag{3.16} \\
&X_{0}^{3}+iX_{0}^{4} =-i\frac{\Phi _{10}\overline{\Psi }_{20}}{(S_{1}-%
\widetilde{S}_{2})_{\xi }}=-i\frac{\Psi _{10}\overline{\Phi }_{20}}{(S_{1}-%
\widetilde{S}_{2})_{\overline{\xi }}}. \tag{3.17}
\end{align*}

The last two expressions in the formulae (3.16), (3.17) are equal to each
other due to the equations (3.9)-(3.10) and differential cosequences of
the constraint $S_{1}+S_{2}=\widetilde{S}_{1}+\widetilde{S}_{2}$.


\begin{proposition}
The formulae (3.12)-(3.15) and (3.8)-(3.10), (3.16), (3.17) define a
conformal immersion of highly corrugated (oscillating) surface
with slow modulation into $R^{4}.$The metric of a surface is given by
\begin{equation*}
ds_{0}^{2}=u_{10}u_{20}dzd\overline{z}=\frac{1}{\varepsilon ^{2}}%
u_{10}u_{20}d\xi d\overline{\xi }, \tag{3.18}
\end{equation*}
Gaussian curvature
\begin{equation*}
K_{0}=-\varepsilon ^{2}\frac{2}{u_{10}u_{20}}\left( \log \left(
u_{10}u_{20}\right) \right) _{\xi \overline{\xi }}, \tag{3.19}
\end{equation*}
squared mean curvature
\begin{equation*}
\mathbf{H}_{0}^{2}=4\frac{\left| p_{0}\right| ^{2}}{u_{10}u_{20}},(3.20)
\end{equation*}
and Willmore functional
\begin{equation*}
W_{0}=4\int \int_{G}\left| p_{0}\right| ^{2}dxdy=\frac{1}{\varepsilon ^{2}}%
4\int \int_{G_{\varepsilon }}\left| p_{0}\right| ^{2}d\xi _{1}d\xi
_{2}, \tag{3.21}
\end{equation*}
where $u_{k0}(\xi ,\overline{\xi })=\left| \Psi _{k0}\right| ^{2}+\left|
\Phi _{k0}\right| ^{2},k=1,2$ ,$\xi =\xi _{1}+i\xi _{2}$ and $G_{\varepsilon
}$ is the rescaled domain G ($\xi _{1}=\varepsilon x,\xi _{2}=\varepsilon
y)$.
\end{proposition}

We shall refer to these formulae as the quasiclassical GWR and corresponding
surfaces as quasiclassical surfaces.For such surfaces the metric is
conformal to a smooth one, Gaussian curvature is small (of the order
$\varepsilon ^{2}$) while the squared mean curvature is finite and smooth.
We emphazise that this quasiclassical GWR  corresponds to the lowest order
terms in the expansions (3.1)-(3.4).

In the three dimensional case for which $p=\overline{p},\Psi _{1}=\Psi
_{2,}\Phi _{1}=\Phi _{2}$ one has $S_{1}=\widetilde{S}_{1}=S_{2}=\widetilde{S%
}_{2}.$ Using the formulae (3.11) in this case one concludes \ that the
quasiclasical GWR in $R^{3\text{ }}$is given by the formulae
\begin{align*}
&X^{1}+iX^{2} =-i\frac{\Psi _{10}^{2}}{2S_{\xi }}\exp
(\frac{2iS_{1}}{\varepsilon }), \\
&X^{3} =\frac{1}{\varepsilon }B(\xi ,\overline{\xi }), \\
&B_{\xi }=\Phi _{10}\overline{\Psi }_{10,} \tag{3.22}
\end{align*}
and
\begin{equation*}
S_{1\xi }S_{1\overline{\xi }}=p_{0}^{2} \tag{3.23}
\end{equation*}
The metric, Gaussian curvature , mean curvature and Willmore functional are
given by formulae (3.18)-(3.21) with $u_{10}=u_{20}=2\left| \Psi
_{10}\right| ^{2}$. For more details on the quasiclassical GWR in $R^{3}$
see [31].

\bigskip

\section{Integrable deformations via the dispersionless DS-II hierarchy.}

\setcounter{equation}{0}

Deformations of quasiclassical surfaces described above are given by the
dispersionless limit of the DS-II hierarchy. To get this limit one, as usual
(at the 1+1-dimensional case, see \textit{e.g.} [30]), assumes that the
dependence of all quantities on $t$ is a slow one, \textit{i.e.} $%
p_{0}=p_{0}(\varepsilon z,\varepsilon \overline{z},\varepsilon
t),S_{k}=S_{k}(\varepsilon z,\varepsilon \overline{z},\varepsilon t)$ and so
on. At the first order in $\varepsilon $ equation (2.8) gives ($\tau
=\varepsilon t_{2})$
\begin{align*}
& S_{\tau }+S_{\xi }^{2}+S_{\overline{\xi }}^{2}-(\omega _{10}+\omega _{20})
=0, \\
&\omega _{10\xi } =2\left| p_{0}\right| _{\overline{\xi }}^{2},
\qquad \omega _{20\overline{\xi }}=2\left| p_{0}\right| _{\xi }^{2},
\tag{4.1}
\end{align*}
while from the system (2.9) one gets
\begin{equation*}
\left(
\begin{array}{c}
i(S_{1\tau }+S_{1\overline{\xi }}^{2}-\omega _{10}),-p_{0}(2\widetilde{S}%
_{1\xi }-S_{1\xi }) \\
\overline{p}_{0}(-\widetilde{S}_{1\overline{\xi }}+2S_{1\overline{\xi }}),-i(%
\widetilde{S}_{1\tau }-\widetilde{S}_{1\xi }^{2}+\omega _{20})%
\end{array}%
\right) \left(
\begin{array}{c}
\Psi _{10} \\
\Phi _{10}%
\end{array}%
\right) =0.
\tag{4.2}
\end{equation*}

The system (2.10) give rises to a systems similar to this. The existence of
nontrivial solutions for the system (3.8) and (4.2) implies the following
independent constraints

\begin{align*}
& S_{1\xi }\widetilde{S}_{1\overline{\xi }}-\left| p_{0}\right| ^{2} =0,
\tag{4.3}
\\
& S_{1\tau }+S_{1\xi }^{2}+S_{1\overline{\xi }}^{2}-2S_{1\xi }\widetilde{S}%
_{1\xi }-\omega _{10} =0,
\tag{4.4} \\
&\widetilde{S}_{1\tau }-\widetilde{S}_{1\xi }^{2}-\widetilde{S}_{1\overline{%
\xi }}^{2}+2S_{1\overline{\xi }}\widetilde{S}_{1\overline{\xi }}+\omega
_{20} =0 .
\tag{4.5}
\end{align*}
This system and similar system for $S_{2}$ and $\widetilde{S}_{2}$ define
deformation of a surface induced by the quasiclassical GWR.

The compatibility conditon for the system (4.3)-(4.5) implies the following
equation for $U=\left| p_{0}\right| ^{2}$
\begin{equation*}
U_{\tau }+2(US_{\xi })_{\xi }+2(US_{\overline{\xi }})_{\overline{\xi }}=0
\tag{4.6}
\end{equation*}%
The difference of equations (4.4) and (4.5) coincides with  equation (4.1)
which can be written also as
\begin{align*}
& S_{\tau }+S_{\xi }^{2}+S_{\overline{\xi }}^{2}+V =0, \\
& V_{\xi \overline{\xi }}+2U_{\xi \xi }+2U_{\overline{\xi }\overline{\xi }}
=0
\tag{4.7}
\end{align*}%
where $V=-(\omega _{10}+\omega _{20}).$ We will refer to the system (4.6),
(4.7) as the dispersionless DS-II equation. It is the 2+1 dimensional
integrable extension of the dispersionless nonlinear Schrodinger equation
studied in [30].

Equation (4.6) implies that for surfaces with rapidly decreasing U as $%
\left| \xi \right| \rightarrow 0$ or for compact surfaces

\begin{equation}
W_{0\tau }=\frac{\partial }{\partial \tau }\left( \frac{1}{\varepsilon ^{2}}%
\int \int_{G_{\varepsilon }}Ud\xi d\overline{\xi }\right) =0 \tag{4.8}
\end{equation}
Thus , deformation of quasiclasical surfaces via the dispersionless DS-II
equation preserves the value of the Willmore functional (3.21).

Similarly, the dispersionless DS-II hierarchy which can be constructed in a
same manner defines integrable deformations of quasiclassical surfaces
generated by quasiclassical GWR. The Willmore functional (3.21) remains
invariant under all these deformations.

For surfaces in $R^{3}$ generated by quasiclassical GWR (3.22), (3.23)
integrable deformations are induced by the dispersionless mVN hierarchy the
lowest member of which is given by the dispersionless limit of the equation
(2.11), \textit{i.e.} by the equation
\begin{align*}
&p_{0\tau }+3\omega _{0}p_{0\xi }+3\overline{\omega }_{0}p_{0\overline{\xi }}+%
\frac{3}{2}p_{0}\omega _{0\xi }+\frac{3}{2}p_{0}\overline{\omega }_{0%
\overline{\xi }} =0,  \\
& \omega _{0\overline{\xi }} =(p_{0}^{2})_{\xi }.
\tag{4.9}
\end{align*}
For more details on this case see [31].

\section{ GWRs in other four dimensional spaces and dispersionless DS system}

GWRs for surfaces and time-like surfaces in $R^{2,2}$ and Minkovsky
space $R^{1,3}$ are rather similar to that in $R^{4}$ [7]. They are based on
the general Dirac system
\begin{align*}
&\Psi _{x} =p\Phi , \\
&\Phi _{y} =q\Psi \tag{5.1}
\end{align*}%
where all variables are complex-valued \ and in concrete cases x and y are
real or complex conjugated and special contraints of the type $q=\overline{p}
$ or $p=\overline{p},q=\overline{q}$ are imposed. Deformations are defined
by the well-known generic DS hierarchy ( two component
Kadomtsev-Petviashvili (KP) hierarchy). Its lowest member is given by the DS
system (see \textit{e.g.} [26,27])

\begin{align*}
&\alpha p_{t} =p_{xx}+p_{yy}+Vp, \\
&\alpha q_{t} = -q_{xx}-q_{yy}-Vq, \\
&V_{xy}+2\left( pq\right) _{xx}+2\left( pq\right) _{yy} =0
\tag{5.2}
\end{align*}
where $\alpha $ is a parameter. This system is equivalent to the
compatibility condition for the system (5.1) and the system
\begin{align*}
&\alpha \Psi _{t} =\Psi _{yy}+\omega _{1}\Psi +p_{x}\Phi -p\Phi _{x}, \\
&\alpha \Phi _{t} =-q_{y}\Psi +q\Psi _{y}-\Phi _{xx}-\omega _{2}\Phi ,
\tag{5.3}
\end{align*}
where $\omega _{1x}=-2\left( pq\right) _{y},\omega _{2y}=-2\left( pq\right)
_{x},V=\omega _{1}+\omega _{2}.$

The quasiclassical version of these formulae and corresponding
quasiclassical GWRs are readely  obtained similar to the previous sections.
Here we will consider only the dispersionless limit of the system (5.2)
since it is of interest also of its own.

First, we introduce the slow variables $\xi =\varepsilon x,\eta =\varepsilon
y,\tau =\varepsilon t$ where $\varepsilon $ is a small parameter and assume
that in the lowest order in $\varepsilon $

\begin{align}
&\Psi =\Psi _{0}\exp \left( \frac{S_{1}}{\varepsilon }\right) , &
\Phi& =\Phi_{0}\exp \left( \frac{S_{2}}{\varepsilon }\right) , \notag \\
& p =p_{0}\exp \left( \frac{S_{1}-S_{2}}{\varepsilon }\right) ,
&  q&=q_{0}\exp
\left( \frac{S_{2}-S_{1}}{\varepsilon }\right)  \tag{5.4}
\end{align}
where $\Psi _{0,}\Phi _{0},p_{0},q_{0}$ are smooth functions of slow
variables. Substituting these expressions into the linear problems (5.1) and
(5.3), one, in the zero order in $\varepsilon $, gets the system of four
equations
\begin{align*}
& S_{1\xi }\Psi _{0}-p_{0}\Phi _{0} =0, \\
& q_{0}\Psi _{0}-S_{2\eta }\Phi _{0} =0 \tag{5.5}
\end{align*}
\begin{align*}
&\left( \alpha S_{1\tau }-S_{1\eta }^{2}-\omega _{10}\right) \Psi _{0}-\left(
S_{1\xi }-2S_{2\xi }\right) p_{0}\Phi _{0} =0, \\
&\left( S_{2\eta }-2S_{1\eta }\right) q_{0}\Psi _{0}+\left( \alpha S_{2\tau
}+S_{2\xi }^{2}+\omega _{20}\right) \Phi _{0} =0 \tag{5.6}
\end{align*}

Equating to zero determinants for all subsystems of the system of equations
(5.5),(5.6), composed by any two equations from them, or simply eliminating
$\Psi _{0}$ and $\Phi _{0}$ from this system , one gets the following system
of three independent equations
\begin{align*}
& S_{1\xi }S_{2\eta }-p_{0}q_{0} =0, \\
&\alpha S_{1\tau }-S_{1\xi }^{2}-S_{1\eta }^{2}+2S_{1\xi }S_{2\eta }-
\omega_{10} =0, \\
& \alpha S_{2\tau }+S_{2\xi }^{2}+S_{2\eta }^{2}-2S_{1\eta }S_{2\eta }+\omega
_{20} =0.\tag{5.7}
\end{align*}
The compatibility condition for these equations is equivalent to the equation
\begin{equation*}
\alpha U_{0\tau }-2\mathbf{\nabla (}U_{0}\mathbf{\nabla }S\mathbf{)=}0
\tag{5.8}
\end{equation*}
where $U_{0}=p_{0}q_{0}$,$S=S_{1}-S_{2}$ and $\mathbf{\nabla }=\left(
\partial _{\xi },\partial _{\eta }\right) $ while the difference of the
second and third equations (5.7) gives
\begin{equation*}
\alpha S_{\tau }-\left( \mathbf{\nabla }S\right) ^{2}+V_{0}=0
\tag{5.9}
\end{equation*}
where
\begin{equation*}
V_{0\xi \eta }-2\triangle U_{0}=0
\tag{5.10}
\end{equation*}
and $\triangle =\partial _{\xi }^{2}+\partial _{\eta }^{2}$.
The system (5.8)-(5.10) represents the dispersionless limit of the DS system (5.2).
We note that the equations (5.9), (5.10) are just the dispersionless limit of
equations (5.2) while equation (5.8) is the dispersionless limit of the
first conservation law \ $\alpha \left( pq\right) _{t}+\left(
pq_{x}-p_{x}q\right) _{x}+\left( pq_{y}-p_{y}q\right) _{y}=0$ for the DS
system (5.2). At $\alpha =-i$ and $q_{0}=-\overline{p}_{0}$ this system
coincides with the system (4.6),(4.7).

The system (5.8)-(5.10) is quite close in the form to the \ classical
hydrodynamical equations of shalow water for gradient flows
(see \textit{e.g.} [30]).
In the one-dimensional limit $\partial _{\xi }=\partial _{\eta }$
it coincides with the one-dimensional Benney system (see [30]). So the
system (5.8)-(5.10) is its two-dimensional integrable generalization.

It differs from the 2+1-dimensional extension of the Benney system,
namely
\begin{align*}
& a_{\tau }+\left( au\right) _{\xi } =0, \\
& u_{\tau }+\frac{1}{2}\left( u^{2}\right) _{\xi }+\omega _{\xi } =0, \\
& \omega _{\eta }+a_{\xi } =0
\tag{5.11}
\end{align*}
given in [32,33] which is equivalent to the compatibility condition for
the system of Hamilton-Jacobi equations
\begin{align*}
&\chi _{\eta } =\frac{a}{\chi _{\xi }-u}, \\
&\chi _{\tau } =-\frac{1}{2}\chi _{\xi }^{2}-\omega .
\tag{5.12}
\end{align*}

To compare this system with (5.7) it is instructive to rewrite (5.7) in the
equivalent form introducing $S$ and $\chi $ defined by $S_{1}=S+\chi
,S_{2}=\chi .$ In these variables the first and second equations (5.7) take
the form
\begin{align*}
&\chi _{\eta } =-\frac{U_{0}}{\chi _{\xi }+S_{\xi }}, \\
&\chi _{\tau } =\chi _{\xi }^{2}-\chi _{\eta }^{2}-2S_{\eta }
\chi _{\eta}-\omega _{20}
\tag{5.13}
\end{align*}

Though the only difference between (5.12) and (5.13) is in their time parts,
the system (5.8)-(5.10) is symmetric in $\xi $ and $\eta $ in contrast to
the system (5.11).

The dispersionless limit of the multi-component KP hierarchy has been
discussed recently in [34]. But, it seems, that the system (5.8)-(5.10) was
missed there.

Finally, it is worth to note that the system (5.8)-(5.10) has a natural
interpretation within the classical mechanics as the system which describes
the integrable deformations of the potential $V_{0}$ in the Hamilton-Jacobi
equation (5.9) driven by equations (5.7), (5.10).

\section*{References}

\begin{enumerate}
\item R. Osserman, Survey of minimal surfaces, Dover Publ., New York, 1986.

\item J.C.C. Nitsche, Lectures on minimal surfaces, Cambridge Univ.Press,
Cambridge, 1989.

\item Global properties of minimal surfaces, Ed. D. Hoffman, Clay
Math.Proc.2, AMS, Providence, RI, 2005

\item B.G. Konopelchenko, Multidimensional integrable systems and dynamics of
surfaces in space, in Proc.'' National workshop on Nonlinear Dynamics''
,(M.Costato, A.Degasperis and M.Milani, Eds.), pp.33-40, Ital. Phys.
Soc.,Bologna, 1995: Preprint of Institute of Mathematics,Academia Sinica,
Taipei, Taiwan, August 1993.

\item B.G. Konopelchenko, Induced surfaces and their integrable dynamics,
Stud.Appl.Math., 96 (1996) 9-51.

\item B.G. Konopelchenko, Weierstrass representations for surfaces in 4D
spaces and their integrable dynamics via the DS hierarchy, Ann.Global
Anal.Geom., 18 (2000) 61-74; arXiv:math/9807129, 1998.

\item B.G. Konopelchenko and G. Landolfi, Induced surfaces and their
integrable dynamics II.Generalized Weierstrass representations in 4D spaces
and deformations via DS hierarchy, Stud.Appl.Math., 104 (1999)
129-169;arXiv:math/9810138, 1998.

\item F. Pedit and U. Pinkall, Quaternionic analysis on Riemann surfaces and
differential geometry of surfaces, Doc. Math.-Extra Volume ICM 1998, II:
389-400.

\item B.G. Konopelchenko and I. Taimanov, \ Generalized Weierstrass formulae,
soliton equations and Willmore surfaces I. Tori of revolution and mKdV
equation, preprint N.187, Univ.Bochum, \ 1995::arXiv:dg-ga/9506011, 1995.

\item R. Carroll and B. Konopelchenko, Generalized Weierstrass-Enneper
inducing, conformal immersions and gravity, Int.J.Mod.Phyd.A, 11 (1996)
1183-1216.

\item I. Taimanov, Modified Novikov-Veselov equation and differential
geometry, Trans. Amer.Math. Soc., Ser. 2, 179 (1997), 133-159.

\item B.G. Konopelchenko and G. Landolfi, Quantum effects for extrinsic
geometry of strings via the generalized Weierstrass representation,
Phys.Lett.B, 444 (1998) 299-308.

\item P. Bracken and M. Grundland, Solutions of the generalized Weierstrass
representation in four dimensional Euclidean space, \ J.Nonlinear Math.Phys.
9 (2002) 357-381.

\item G. Landolfi, New results on the Canham-Helfrich membrane model via the
generalized Weierstrass representation, J.Phys.A: Math. Gen., 36 (2003)
11937-11954.

\item I. Taimanov, The two-dimensional Dirac operator and the theory of
surfaces, Russian Math.Surveys, 61 (2006) 79-159.

\item Jianhua Chen and Weihuan Chen, Generalized Weierstrass representation
of surfaces in $R^{4}$, J.Geom.Phys., 57 (2007) 367-378.

\item S. Matsutani, Generalized Weierstrass relations and Frobenius
reciprocity, Math. Phys. Anal .Geom.  (2007).

\item E. Feinberg, On the propagation of radio waves along an imperfect
surface, Acad. Sci. USSR. J. Phys., \textbf{8} (1944), 317-330.

\item R.A. Hurd, The propagation of an electromagnetic wave along an infinite
corrugated surface, Canad. J. Phys., \textbf{32 }(1954), 727-734.

\item S. Asano, Reflection and refraction of elastic waves at a corrugated
boundary surface, Bull. Earthquake Res. Inst. Tokyo, I,\textbf{\ 38} (1960),
177-197: II, \textbf{39 }(1961), 367-466.

\item I.H. Sabitov, The rigidity of corrugated surfaces of revolution, Mat.
Zametki, \textbf{34 }(1973), 517-522.

\item J. Krug and P. Meakin, Kinetic roughening of Laplacian fronts, Phys.
Rev. Lett., \textbf{66 }(1991), 703-706.

\item D.A. Lidar, Inversion of randomly corrugated surface structure from
atom scattering data, Inverse Problems, \textbf{14 }(1998), 1299-1310.

\item V. Jaksic, Spectral theory of corrugated surfaces, Journees'' Equations
aux Derives Part. '', \textbf{VIII }\ (2001), 11.

\item Y. Tsori and D. Andelman, Parallel and perpendicular lamellae on
corrugated surfaces, Macromolecules, \textbf{36 }(2003), 8560-8573.

\item M.J. Ablowitz and P. Clarkson, Solitons, nonlinear evolution equations
and inverse scattering, Cambridge Uni.Press, Cambridge, 1991.

\item B.G. Konopelchenko, Introduction to multidimensional integrable
equations, Plenum Press, New York, 1992.

\item G.B. Whitham, Linear and nonlinear waves, JohnWiley \& Sons, New York,
1974.

\item V.P. Maslov, Complex WKB method in nonlinear equations, Nauka, Moscow,
1977.

\item V.E. Zakharov, Benney equations and quasiclasical approximation in the
method of the inverse problem, Func.Anal.Appl., 14 (1980) 89-98.

\item B.G. Konopelchenko, Corrugated surfaces with slow modulation and
quasiclassical Weierstrass representation, arXiv:nlin.SI/0606004, 2006.

\item V.E. Zakharov, Dispersionless limit of integrable systems in 2+1
dimensions, in: Singular Limits of Dispersive Waves,
(Eds. N.M.Ercolani et al.), Plenum Press, New York, 1994.

\item I.M. Krichever, The tau-function of the universal Whitham hierarchy ,
matrix models and topological field theory, Commun. Pure Appl. Math., 47
(1994) 437-475.

\item K. Takasaki and T. Takebe, Universal Whitham hierarchy, dispersionless
Hirota equations and multi-component KP hierarchy, arXiv:nlin/0608068, 2006.
\end{enumerate}

\end{document}